\documentclass{cjaa}    
\usepackage{graphicx}
\usepackage{natbib}

\newcommand\sgn{\ensuremath {\mathrm{sgn}\,}}

\begin{document}
\title{Polarised views of the drifting subpulse phenomenon}
\volnopage{Vol.0 (2006) No.0, 000--000}
\setcounter{page}{1} 
\author{Russell T. Edwards}
\institute{Australia Telescope National Facility, CSIRO, P.O. Box 76, Epping NSW 1710 Australia\\
\email{Russell.Edwards@csiro.au}}
\date{Received}

\abstract{I review recent results concerning the shape of drifting
subpulse patterns, and the relationship to model predictions. While a
variety of theoretical models exist for drifting subpulses, observers
typically think in terms of a spatio-temporal model of circulating
beamlets. Assuming the model is correct, geometric parameters have
been inferred and animated ``maps'' of the beam have been made. However,
the model makes very specific predictions about the curvature of the
drift bands that have remained largely untested. Work so far in this
area indicates that drift bands tend not to follow the prediction, and
in some cases discontinuities are seen that are suggestive of the superposition
of out of phase drift patterns. Recent polarimetric observations also
show that the drift patterns in the two orthogonal polarisation modes
are offset in phase. In one case the pattern in one of the modes shows
a discontinuity suggesting no less than three superposed, out-of-phase
drift patterns! I advise caution in the interpretation of observational
data in the context of overly simplistic models.
\keywords{pulsars: general}
}

\maketitle

\section{The Carousel Model}
\begin{figure}
\centering\resizebox{7cm}{!}{\includegraphics{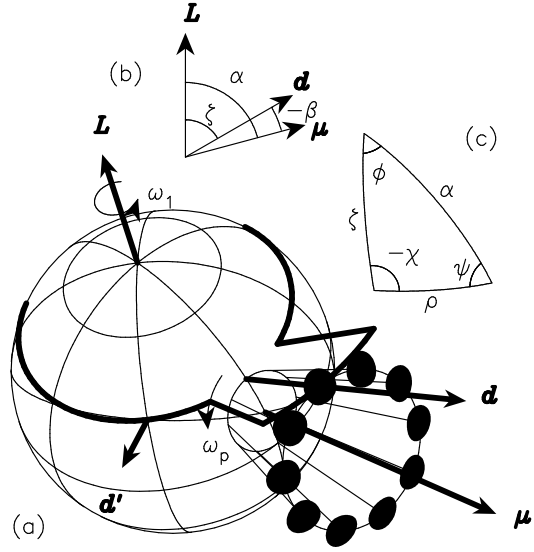}}
\caption{Diagram of sub-pulse emission geometry, reproduced from
\citet{es02}. In part (a) the angular momentum, magnetic moment, and
line-of-sight vectors are shown with symbols {\boldmath $L$, $\mu$,
and $d$} respectively. Part (b) shows the main vectors as they appear
in the plane they share when the sight-line makes its closest approach
to the magnetic pole. The angles $\zeta$ (between the sight-line and
the spin axis), $\alpha$ (between the magnetic and spin axes) and
$\beta\equiv\zeta-\alpha$ (between the sight-line and the magnetic
axis at their closest approach) are shown.  In part (a) a second line
of sight ($\vec{d'}$) is drawn, along with the meridian it shares with
with the magnetic pole. The spherical triangle formed by $\vec{L}$,
$\vec{\mu}$ and $\vec{d'}$ is duplicated in part (c).}
\label{fig:schematic}
\end{figure}

Although the drifting subpulses phenomenon has now been approached from
a number of different theoretical perspectives
(e.g. \citealt{zhe71,kmm91,wri03,cr04a,you04,gmml05,fkk06}), most
observational studies have assumed the applicability of the so-called
carousel model \citep{rud72}. This model postulates that the modulations
have a spatio-temporal origin: the subpulse structure is due to the
passing of the line of sight over multiple ``beamlets'', while the
temporal, pulse-to-pulse modulation is due to the slow circulation of the
beamlet system (Fig.\ \ref{fig:schematic}(a)). In the usual physical
model, the beamlets consist of radiation beamed tangentially to local
magnetic field lines, within ``tubes'' of plasma, each flowing outward
from a localised breakdown (``spark'') of a potential gap over the
magnetic pole \citep{rud72}. The circulation of the sparks is attributed
to {\boldmath $\vec{E} \times \vec{B}$} drift. 

\citet{es02} showed that the carousel model makes very specific predictions
regarding the shape of drift bands. Specifically, assuming that the
beamlets are uniformly spaced in magnetic azimuth, the variation of
subpulse phase with pulse longitude obeys a geometric relationship. This
is because it is linearly tied via the number of beamlets ($N$) to the
magnetic azimuth of the observer ($\psi$), which in turn obeys the following
relation:
\begin{equation}
\tan{\psi} = \frac{\sin\phi\sin\zeta}
        {\cos\zeta\sin\alpha - \cos\phi\sin\zeta\cos\alpha} ,
\label{eq:psiphi}
\end{equation}
\noindent where $\phi$ is pulse longitude, $\zeta$ is the angle
between the spin axis and the line of sight, and $\alpha$ is the angle
between the magnetic and spin axes (see Fig.\
\ref{fig:schematic}(b)). The reader may recognise the similarity
between this relation and the common ``rotating vector model'' for the
position angle of linear polarisation, $\chi$. The reason for this
similarity is that all of the pertinent angles are related according
to the spherical triangle of Fig.\ \ref{fig:schematic}(c). The
subpulse phase, $\theta$, is linearly related to magnetic azimuth, plus
a generally small additional term describing the circulation of the
carousel over the course of the pulse:
\begin{eqnarray}
\theta(\phi) &=&
        -N \sgn\beta \tan^{-1} \left[ 
        \frac{\sin\phi\sin\zeta}
        {\cos\zeta\sin\alpha - \cos\phi\sin\zeta\cos\alpha}
        \right] 
     + \phi \left(n + \frac{P_1}{\hat{P_3}}\right) + \theta'.
\label{eq:thetaphi}
\end{eqnarray}
\noindent Here $\theta'$ is an arbitrary constant, $P_1$ is the pulse
period, $\hat{P_3}$ is the observed pulse-to-pulse periodicity of the
drift bands (including a sign which specifies the direction of the
drift slope), and $n$ is an integer specifying the order of aliasing
present.

It is important to note that by examining the dependence of modulation
phase upon pulse longitude, this method tracks the drift band peaks
according to local maxima along lines of constant longitude. This is
in contrast to the more common technique of finding local maxima in
individual pulses (i.e. lines of constant pulse number). The former
method offers an important advantage: the amplitude-phase
decomposition conveniently separates beamlet azimuthal spacing
(i.e.\ circulation or subpulse phase) from colatitudinal amplitude
windowing (i.e. subpulse amplitude and mean profile shape). In the
latter method, subpulse peaks are ``pushed'' in pulse longitude, in
the direction of increasing subpulse amplitude, and therefore do not
cleanly track the azimuthal spacing of beamlets (see \citealt{es02} and
\citealt{es03b}).

\section{Why Test the Carousel Model?}
While the carousel model has been in use for over two decades in the
interpretation of drifting subpulse patterns, until recently very
little use had been made of the opportunity presented to test the
model (however, see \citealt{wri81}). Many examples exist of
inferences being made on the basis of the correctness of the model,
without any checks being made of the validity of this assumption. For
brevity I will focus upon a critical look at one particularly
interesting series of studies, namely the industry of ``polar cap
mapping'' of systems of drifting subpulses
\citep{dr99,dr01,ad01,rsd03,ad05}.

\citet{dr99} (hereafter DR99) detected, for the first time, a pair of
``outrider'' components surrounding the first harmonic of the drifting
subpulse pattern in the fluctuation spectrum of PSR B0943+10. While
relatively marginal in terms of signal to noise ratio, and only
detected in a small section of one of several observations, the
detection was nevertheless exciting since it was potentially
supporting evidence for the carousel model.  This is because under
that model, any overall overall pattern of intensity around the ring
of beamlets would result in the convolution of all Fourier components
with the Fourier transform of the azimuthal intensity dependence
\citep{es02}. For an intensity pattern consisting of a strong steady
component plus a weak variation $\propto \sin(\psi-\psi_0)$, this adds
a small pair of components spaced $1/\hat{P_3}$ from the parent
component. Although no such ``outriders'' appear around the DC
component, DR99 pursued the possibility that this was the explanation
for the extra components around the first harmonic.

Having assumed the applicability of the carousel model, DR99 set out
to establish the values of all geometric parameters necessary to
invert the transformation between the magnetic frame and the
observer's frame, in order to ``map'' the observed intensity onto an
animated display of the beam intensity distribution (or, equivalently,
the polar cap excitation distribution). The first hurdle in this
process is the unknown aliasing order, $n$. The traditional method of
fluctuation spectrum analysis, the longitude-resolved fluctuation
(LRFS), is insensitive to the sign or direction of drift, thereby
doubling the number of aliasing possibilities in Eq.\
\ref{eq:thetaphi}. DR99 used a new technique called the
harmonic-resolved fluctuation spectrum, later shown to be equivalent
to the two-dimensional fourier transform of the longitude-time
dependence of intensity \citep{es02,esv03}. Having eliminated the
aliasing possibilities previously admitted by the LRFS, DR99
apparently considered the aliasing question solved. However, even
using the improved frequency measurements of \citet{dr01}, an infinite
range of possible values of $n$ remain \citep{es02}.
 
To obtain the other geometric parameters, \citet{dr01} turned to
polarimetric observations. Although the rotating vector model has
been shown to hold in only a minority of pulsars \citep{ew01}, and
B0943+10 exhibits a featureless linear sweep of position angle giving
no clear evidence for applicability of the model, it was taken by
\citet{dr01} to apply. The values derived were reported to support
the chosen aliasing solution, although as noted above other solutions
are possible and only by combining the spectral results and with
polarimetry can a unique solution be obtained \citep{es02}.

Having obtained nominal values for all relevant geometrical
parameters, \citet{dr01} proceeded to form an animated map of the
emission beam, a product which, if reliable, offers remarkable
insights into conditions on the polar cap and/or in the
magnetosphere. However, in my view this result is far from water
tight. It relies on the reality of certain spectral features of
marginal significance, the applicability of the rotating vector model
for polarisation, and the applicability of the geometric predictions
of the carousel model for drifting subpulses.  Of the latter two
conditions, the first has been proven to be generally false, and the
second was until recently almost entirely untested. This was seen a
sufficient motivation to rigorously test the geometrical predictions
of the model. 

As I will outline below, the carousel model in its basic form has now
been shown to be generally inapplicable. It is therefore advisable to
approach any inferences made from pulsar data on this basis, including
polar cap maps, with caution.

\section{Testing the Carousel Model}
\begin{figure}
\begin{center}
\begin{tabular}{lr}
\resizebox{4.5cm}{!}{\includegraphics{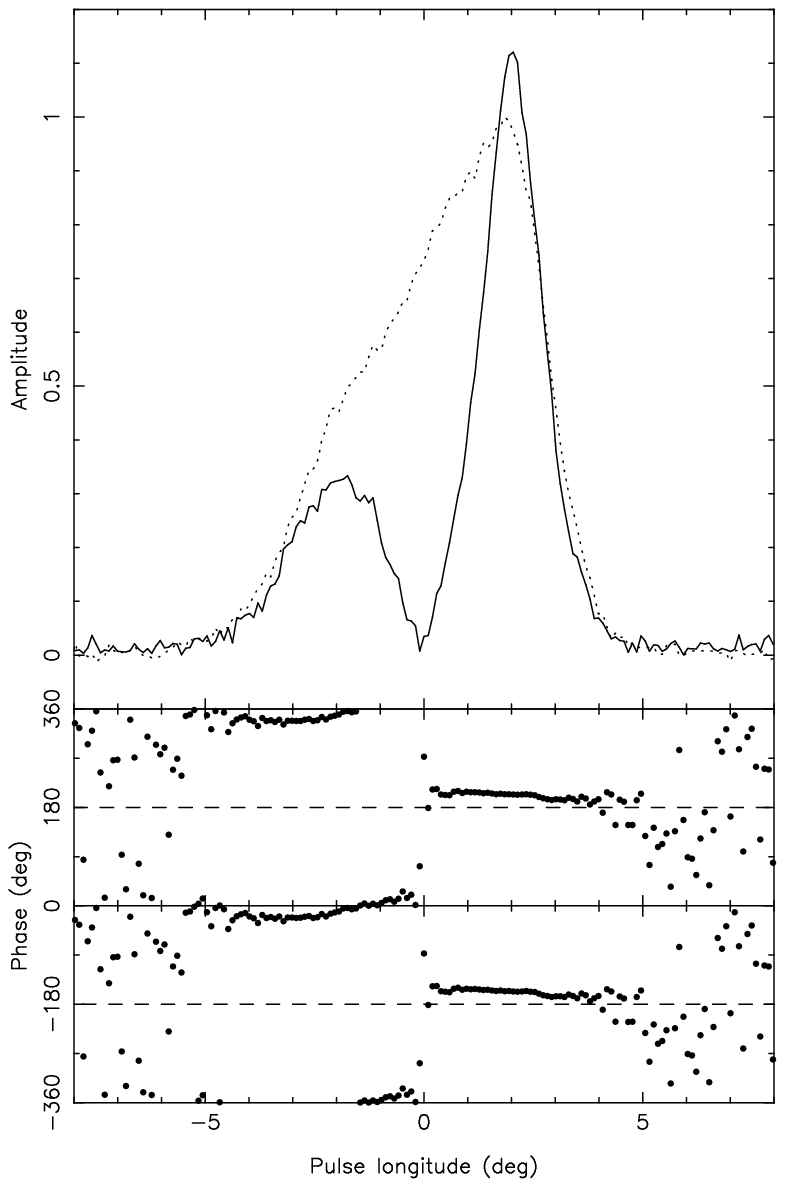}} &
\resizebox{4.0cm}{!}{\includegraphics{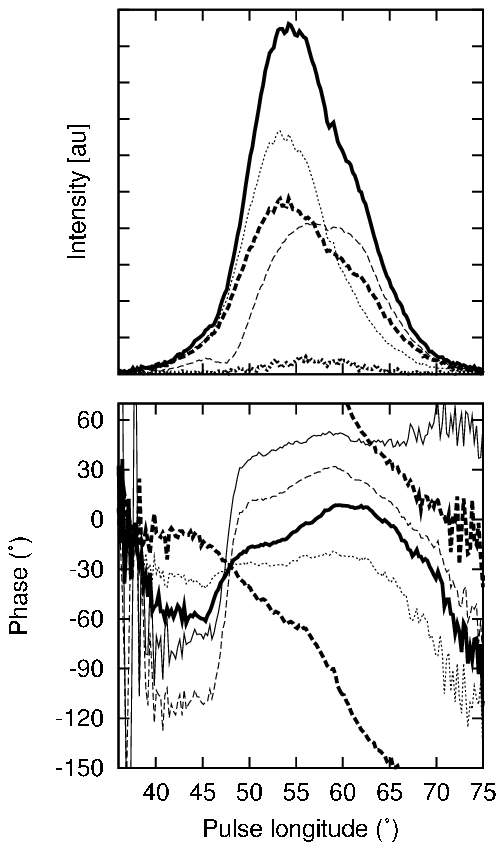}}
\end{tabular}
\end{center}
\caption{Subpulse phase jumps. (Left) Pulse profile (dotted line),
subpulse amplitude (solid line) and phase (points, plotted twice with
linear slope subtracted) for PSR B0320+39 at 328~MHz. (Right) Pulse
profile (thick solid line), subpulse amplitudes and phases for
orthogonal polarisation modes (thin dotted and dashed lines, slope
subtracted), for PSR B0809+74 at 328~MHz (see \citealt{edw04} for full
description).}
\label{fig:phasejumps}
\end{figure}

For reasons outlined in the preceding sections, a study was conducted
of the subpulse phase variation in several pulsars. The first pulsar
to be examined was PSR B0320+39, at a frequency of 328 MHz
\citep{esv03}. Surprisingly, the subpulse phase was shown to exhibit a
phase ``jump'' of roughly 180\degr, near the centre of the profile
(Fig.\ \ref{fig:phasejumps}). Such a feature cannot be reconciled with
the predictions of the carousel model in its simplest form. Several
features of the drift bands are suggestive of an origin in superposed,
out-of-phase patterns:
\begin{itemize}
\item the rapid 180\degr\ phase jump,
\item the vanishing subpulse amplitude at this point,
\item the continuity of the absolute value of the rate of change of subpulse amplitude at this point, and
\item the continuity of, and indeed, absence of any feature whatsoever
in the total intensity profile at this point.
\end{itemize}
\noindent Such features are familiar from other cases of superposed wave phenomena,
for example in the presence of superposed orthogonal polarisation modes
of shifting dominance.

Two possibilities were discussed by \citet{esv03} for the origin of the
putative ``double images'' of the drifting subpulse pattern, both
invoking an underlying circulating carousel system. The first suggests
that the two images originate from either side of the magnetic pole,
with that from the far side reaching the observer by virtue of
magnetospheric refraction. For an odd number of beamlets and an
axisymmetric plasma distribution, the images will be in antiphase.
The second follows the suggestion of \citet{ran93} that emission
occurs at two discrete heights in the magnetosphere. The divergence of
field lines results in nested cones of emission, while aberration and
retardation shift their centres somewhat. However, the difference in
emission height needed for sufficiently offset cones in this case is
in excess of 10,000 km, which is strongly incompatible with other
estimates of pulsar emission heights (e.g. \citealt{drh04}).

The detection of the phase jump in PSR B0320+39 also led to the
development of a completely new model for drifting
subpulses. \citet{cr04a} suggested that drifting subpulses are a
manifestation of non-radial oscillations of high wavenumber. In this
model, the phase jump and amplitude nulling are neatly explained as
corresponding to the passage of the line of sight over a nodal point
in the oscillation. However, as a temporal oscillation, the subpulse
phase should be a strictly linear function of pulse longitude, which
is incompatible with minor but definite deviations from linearity
detected by \citet{esv03}. Although amplitude windowing of the
subpulses may produce apparently curved drift bands \citep{cr04a}, as
noted in Sect. 1, the prime advantage of using subpulse phase to test
models of drifting subpulses is that it is in fact unaffected by such
windowing.

Further studies revealed more complex deviations for the simple model
predictions. \citet{es03b} found that at 1380~MHz, PSR B0320+39 showed
subpulses that were much weaker, with a phase slope that contained
strange deviations from linearity and only weak evidence for a phase
jump. Furthermore, observations of PSR B0809+74 revealed unexplained
deviations from the model curve at 328~MHz, which evolved to include a
rapid phase jump of some $\sim 120$\degr at 1380~MHz, accompanied by
an attenuation of amplitude. Analogous to similar polarimetric
behaviour attributed to non-orthogonal polarisation modes, the latter
was interpreted as suggestive of the superposition of subpulse
patterns that were not in pure antiphase, and potentially were not
offset in phase by a constant amount across whole the pulse longitude
range. \citet{es03b} suggested that such circumstances could also
explain the curious bidirectional subpulse drift seen in PSR B1918+19,
in much the same way that polarisation position angle sweeps can
undergo erratic jumps and sense changes in the presence of non-orthogonal
radiation. Since then, further examples of phase steps
\citep{wes06} and bidirectional drift \citep{clm+05,wes06} have been
found.

Observations made over thirty years ago have shown that the
polarisation at a given pulse longitude is periodically modulated
between two orthogonal states, for at least two pulsars
\citep{mth75}. Following a high resolution study of PSR B0809+74 by
\citet{rrs+02}, it was suggested that this effect could be explained
via the superposition of out-of-phase subpulse patterns in the two
orthogonal modes \citep{esv03,rr03}. By decomposing the signal into
its component orthogonal modes, \citet{edw04} showed that this was
indeed the case in observations made at 328~MHz. Moreover, one of the
modes showed a sharp phase jump accompanied by a reduction in subpulse
amplitude (Fig.\ \ref{fig:phasejumps}), very similar to the feature
seen in total intensity at 1380~MHz.  The other mode showed no feature
at all in this region, supporting the physical validity of the
orthogonal mode decomposition.  The implications of this result are
startling: if the orthogonal modes are out of phase due to
superposition then there are at least two images, but if the phase
jump in one of the modes is also due to out-of-phase superposition,
then there must be at least three distinct, out of phase ``images'' of
the subpulse pattern, with polarisation effects somehow tied in to the
multiple imaging process.

Single pulse polarimetry of PSR B0809+74 at 1380 MHz, and observations
of two other pulsars at lower frequencies revealed significantly more
complex behaviour. In these observations, the polarisation vector
(composed of Stokes Q, U and V) was found to move along a periodic locus
at the frequency of the subpulse modulation, however the first harmonic of
this locus was elliptical, rather than linear as expected under the 
superposition of orthogonal polarisation modes. 

\section{Summary and Conclusions}
Subpulse phase analysis presents a sensitive means for testing strong
geometric predictions made by the standard carousel model for drifting
subpulses. Testing this model is important because a variety of
inferences have been made from observational results under the
assumption that the model applies. In particular, procedure of
``mapping'' the polar cap of PSR B0943+10 relies heavily on this
model, as well as assuming, despite the odds, that the geometric model
for polarisation applies to this pulsar. A number of studies of
subpulse phase have now been conducted, with results that strongly
disagree with the standard model. Several features of the observed
phase slopes are suggestive of the superposition of out of phase
``images'' of subpulse patterns. A polarimetric study of PSR B0809+74
indicates that out of phase, orthogonally polarised subpulse patterns
are observed in superposition, and that these patterns themselves can
consist of the superposition of two or more out of phase
components. At other frequencies, complex, non-orthogonal periodic
modulations of the polarisation were seen, with unknown origin.

In light of the strong deviations from the standard model observed in
all cases so far, I suggest caution be exercised in the interpretation
of observational results that assume its validity.


\end{document}